\documentclass[12pt,english]{article}
\usepackage{mathptmx}
\usepackage{courier}
\usepackage[T1]{fontenc}
\usepackage{geometry}
\usepackage{babel}
\usepackage{booktabs}
\usepackage{textcomp}
\usepackage{amsbsy}
\usepackage{graphicx}
\usepackage{setspace}
\usepackage[authoryear]{natbib}
\usepackage[unicode=true]{hyperref}
\usepackage{setspace}
\usepackage{amsmath}
\usepackage{eurosym}
\usepackage{placeins}

\providecommand{\keywords}[1]
{
  \small	
  \textbf{\textit{Keywords---}} #1
}

\date{\today}

\begin{document}
%\doublespacing

\title{Credit Crunch:
The Role of Household Lending Capacity in the Dutch Housing Boom and Bust 1995-2018}

\author{Menno Schellekens$^{1}$ and Taha Yasseri$^{1,2,3,4}$\footnote{Corresponding author: Taha Yasseri, D405 John Henry Newman Building, University College Dublin, Stillorgan Rd, Belfield, Dublin 4, Ireland. Email: taha.yasseri@ucd.ie.}\\
{\small $^1$Oxford Internet Institute, University of Oxford, Oxford, UK} \\
{\small $^2$School of Sociology, University College Dublin, Dublin, Ireland}\\ 
{\small $^3$Geary Institute for Public Policy, University College Dublin, Dublin, Ireland}\\
{\small $^4$Alan Turing Institute for Data Science and AI, London, UK}}

\maketitle

\begin{abstract}
What causes house prices to rise and fall? Economists identify household access to credit as a crucial factor. "Loan-to-Value" and "Debt-to-GDP" ratios are the standard measures for credit access. However, these measures fail to explain the depth of the Dutch housing bust after the 2009 Financial Crisis. This work is the first to model household lending capacity based on the formulas that Dutch banks use in the mortgage application process. We compare the ability of regression models to forecast housing prices when different measures of credit access are utilised. We show that our measure of household lending capacity is a forward-looking, highly predictive variable that outperforms `Loan-to-Value' and debt ratios in forecasting the Dutch crisis. Sharp declines in lending capacity foreshadow the market deceleration.
\end{abstract}


\section{Introduction}
The flow of credit from the financial sector to the housing market is critical for understanding house prices, because households usually finance properties with mortgage debt. The availability of credit determines how much households can borrow and thus how much they can bid on properties. As relaxed credit constraints allow all market participants to borrow more, households often have to borrow more to stay competitive, and house prices rise quickly \citep{bernanke1995inside,kiyotaki1997credit}. After the Great Financial Crisis, scholars started emphasising the importance of credit conditions to the development of housing prices. 

Studies of housing markets in advanced economies consistently find that empirical models that include measures of credit conditions outperform models that do not.\footnote{Such studies have been conducted for the United States \citep{Duca2016}, Ireland \citep{Lyons2018}, Finland \citep{Oikarinen2009}, Norway \citep{anundsen2013self}, France \citep{chauvin2014consumption} and Sweden \citep{Turk2016}.} Competing measures of credit conditions have emerged in the literature. Some authors employ the average `loan-to-value' (LTV) ratio of first time buyers \citep{Duca2016}.\footnote{The LTV is the mortgage for a property divided by the price paid. A low LTV means that the lender left a large margin of safety between the mortgage and the market value of the home. Thus, the value of the collateral would be greater than the mortgage even if the house decreases in value. A high LTV indicates that lenders are willing to tolerate more risk.} Others choose mortgage debt to GDP ratios \citep{Oikarinen2009}, survey data from senior bank employees \citep{vanderVeer2016} and indeces of the `ease' of credit policy \citep{chauvin2014consumption}. 

\begin{figure}[!htb]
  \includegraphics[width=.33\linewidth]{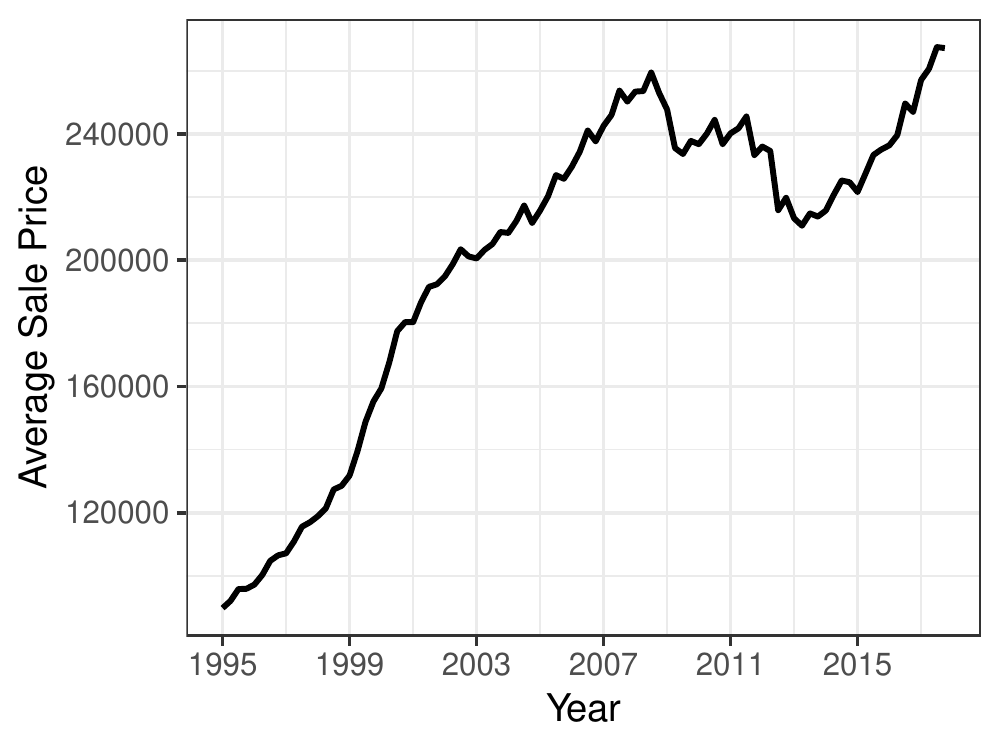}
  \includegraphics[width=.33\linewidth]{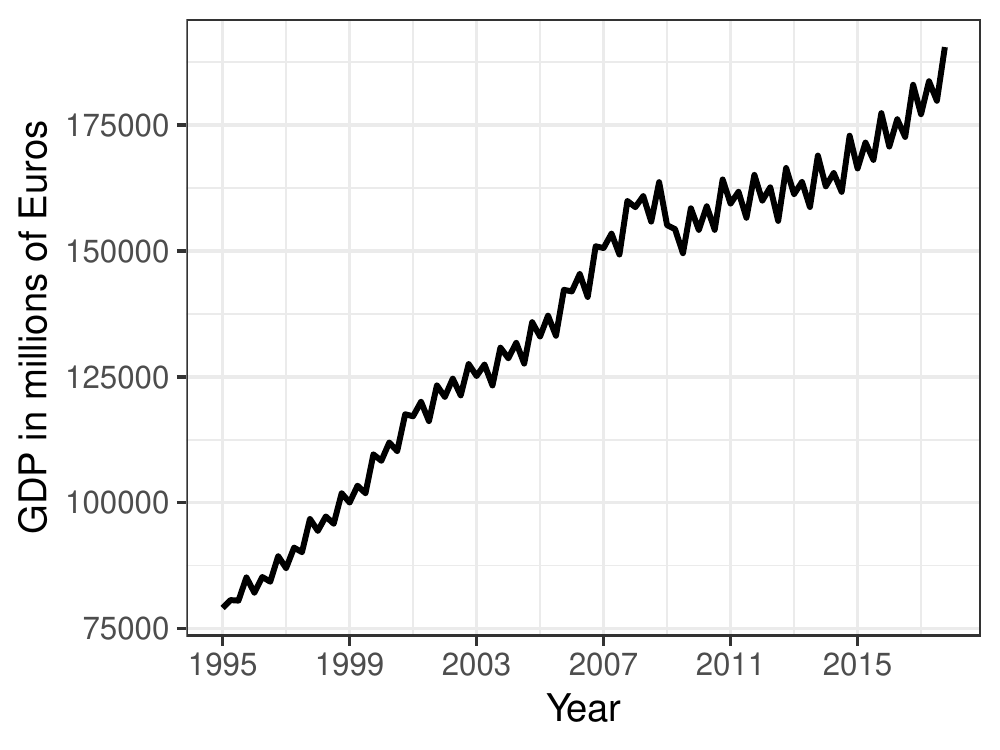}
  \includegraphics[width=.33\linewidth]{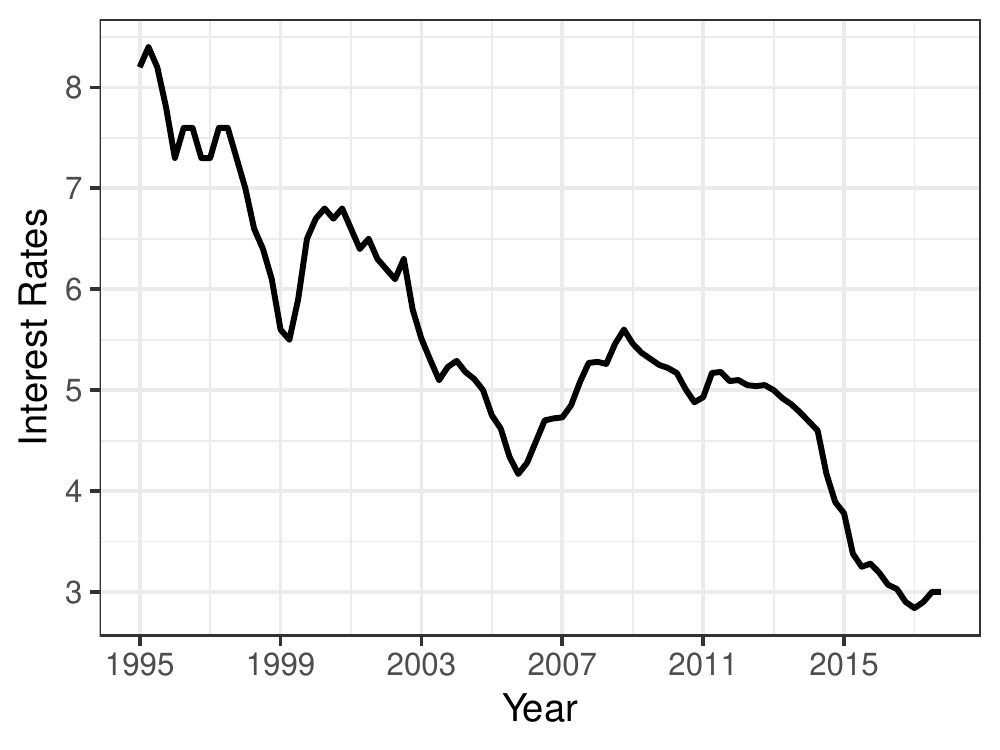}
\caption{ Statistics From The Central Bureau of Statistics (CBS) in the Netherlands 1995-2018 \label{fig:NedStats}}
\end{figure}

The Dutch housing market experienced a boom and bust in the period 1995-2018. The Dutch case is a puzzle because existing models do not explain the size of the boom and the depth of the bust. Figure \ref{fig:NedStats} shows that house prices rose rapidly in the 1990's and early 2000's and entered a period of sustained decline after 2009. Despite falling interest rates, housing prices fell 16\% from their peak value. The crisis came at great cost to many Dutch families. In 2015, 28\% of homes were deemed `under water': the home was worth less than the outstanding mortgage debt \citep{CPB2014}. 

Based on a qualitative analysis of Dutch housing reforms, we develop a novel approach to forecasting house prices that utilises a new measure of credit access. We study the `Loan-to-Income' formulas that banks use to calculate how much money households can borrow relative to their income. We model these formulas and calculate lending capacity for the average Dutch household from 1995 to 2018 with three parameters: average household income, mortgage interest rates and regulatory changes. We find household lending capacity is a more accurately predictor of house prices in the Netherlands than LTV ratios and `debt-to-GDP' ratios. In a test on out-of-sample data, a univariate OLS model with household lending capacity provides the most robust forecasts.

The next section provides an analysis of the formulas that govern access to credit in the Netherlands and how we model these formulas. In the methods section, we describe our dataset and the specification of statistical models. Lastly, we report our findings and discuss the implications and limitations of our approach.

\section{Modelling Household Lending Capacity \label{sec:theory}}  
The notion that household lending capacity ($HLC$) - the amount of mortgage debt one can legally borrow to finance a home - influences house prices is not new. Economists generally believe that the ability to afford loans is one of the channels through which income and interest rates affect prices \citep{ESB2017}. We hypothesise that the details of how $HLC$ is calculated matter. The rules that govern household borrowing relative to income are called `Loan to Income' (LTI) formulas. This section is divided in two parts: an introduction to LTI formulas in the Netherlands and an elaboration on how we model LTI formulas.

\subsection{LTI Formulas in the Netherlands}
Before 2009, the Netherlands had no national laws or regulations that restricted lending capacity. However, the associations of banks and insurers - the dominant mortgage providers - authored the `Gedragscode Hypothecaire Financieringen' (Behavioral Code Mortgage Finance, GHF) in 1999. GHF specified norms that the members agreed to adhere to for fair and responsible mortgage finance, including norms pertaining to the height of mortgages. The association specified that households would not be allowed to spend more than a specific percentage of their available income on mortgage repayment, depending on their family type and income bracket \citep{NIBUD2018}. For the average family, the maximum percentage of income designated for income (``woonquote'') was a range of 21-40\% depending on their economic circumstances. Banks were allowed to deviate from these norms in exceptional cases, but the Dutch regulator found that the norms were mostly adhered to \citep{DNBm}. The norms were tightened in response to the Financial Crisis and translated into law in 2011 \citep{RABOm, Rabo2015}.

The housing market faced a number of reforms in the period 2011-2018. % TODO Make list later? Maybe from literature?
One key reform directly impacted lending capacity. After the advent of the Financial Crisis of 2007 - in which mortgage debt played a key role in the American crisis and to a lesser extent the near collapse of Dutch banks - the Dutch government aimed to decrease the burden of mortgage debt on the economy \citep{IMF2019Netherlands}. The Dutch Central Bank identified a category of mortgage products - `krediethypotheken' (interest-only mortgages) - as a source of increasing mortgage debt. Interest-only mortgages grew their market share from 10\% to 50\% in the period 1995-2008 \citep{DNBm}. Customers with interest-only plans could choose to pay off their mortgage at their own pace or not at all, whereas traditional annuity and linear payment plans require households to amortise their mortgage every month. The fact that borrowers only had to budget interest payments meant that households could borrow more with an interest-only mortgage than with a traditional annuity. As of 2011, the government required lenders to use the maximum annuity as the maximum mortgage, even for an interest-only mortgage \citep{Notitie2018, RABOm}. As of 2013, the government no longer allowed households to deduct interest payments from taxes for new interest-only mortgages \citep{DNB2019LTV}. This removed a crucial fiscal benefit from the category and made interest-only mortgages more expensive than annuity mortgages\footnote{The Dutch government decided to gradually phase out the mortgage interest deduction altogether in 2014.}. The next section details how the transition between regulatory regimes is modelled.

\subsection{Modelling LTI formulas}
LTI formulas govern how much a household can borrow given their income and mortgage interest rates. Lenders use formulas to calculate the legal limit that households can borrow based on income, interest rates, risk of default, expenses, debts, housing costs and more. We boil the formulas down into a single equation for annuities and interest-only mortgages.

$$ LC_{k} = \frac{I_{h}}{ r + c } $$

\noindent where $I_{h}$ is annual income allocated to housing expenses, $r$ is annual interest rates and $c$ is a fixed annual cost expressed as a percentage of the value of the home. The effect of interest rate changes on lending capacity is neither linear nor independent of income. This is significant, because independence and linearity are key assumptions in most regression models. This is also true of annuities. The maximum annuity mortgage is the mortgage for which households can pay the yearly interest and mandatory amortisation. The simplified equation for the maximum annuity mortgage $LC_{a}$ reads:

$$ LC_{a} = I_{h} * f( r + c ) $$

\noindent where $I_{h}$ is income allocated to housing expenses, $r$ is interest rates and $c$ is a fixed annual cost expressed as a percentage of the value of the home. Function $f(x)$ is the standard annuity formula:

$$ f(x) = \frac{1 - (1 + \frac{x}{12})^{(-360)})}{\frac{x}{12}} $$ 

Annuity mortgages are calculated differently than interest-only mortgages, so they produce different lending capacities for the same income and interest rates. In summary, the LTI formulas are non-linear functions in which the effect of a change in income and interest rates is dependent on the value of the other variable.

\section{Data \& Methods \label{sec:methods}}
Given that the dependencies and non-linearities cannot be modelled accurately in linear regression models, we construct the measure `average household lending capacity' ($HLC$). The following section has three purposes. First, it lays out the variables and their summary statistics. Second, it details the construction of the key variable $HLC$. Finally, it describes the models and model evaluation metrics.

\subsection{Data \label{sec:data}}
We primarily utilise publicly available income and price data from Dutch Central Bureau of Statistics (CBS) and the `De Nederlandsche Bank` (DNB), the Dutch central bank. Summary statistics can be found in Table \ref{tab:sumstats}; details on sources and variable construction in the Appendix.

% Table created by stargazer v.5.2.2 by Marek Hlavac, Harvard University. E-mail: hlavac at fas.harvard.edu
% Date and time: Fri, Jul 19, 2019 - 11:01:48
\begin{table}[!htbp] \centering 
  \caption{Summary Statistics} 
  \label{tab:sumstats} 
\begin{tabular}{@{\extracolsep{5pt}}lccccccc} 
\\[-1.8ex]\hline 
\hline \\[-1.8ex] 
Statistic & \multicolumn{1}{c}{N} & \multicolumn{1}{c}{Mean} & \multicolumn{1}{c}{St. Dev.} & \multicolumn{1}{c}{Min} & \multicolumn{1}{c}{Pctl(25)} & \multicolumn{1}{c}{Pctl(75)} & \multicolumn{1}{c}{Max} \\ 
\hline \\[-1.8ex] 
Average House Price & 92 & 201.180 & 50,896 & 89.792 & 179.627 & 237.662 & 267.464 \\ 
Household Income & 91 & 68.733 & 13.467 & 44.916 & 59.815 & 77.209 & 98.783 \\ 
Interest Rates & 92 & 5,382 & 1,338 & 2,840 & 4,728 & 6,325 & 8,400 \\ 
Avg. LTV & 88 & 101,3 & 2,6 & 96 & 100,2 & 103,4 & 103,9 \\ 
Interest-Only Marketshare & 92 & 0,217 & 0,181 & 0,000 & 0,000 & 0,410 & 0,463 \\ 
\hline \\[-1.8ex] 
\end{tabular} 
\end{table}

\subsection{Model Specifications \label{sec:specification}} 
Three model are tested: (1) a benchmark model with LTV values as a measure of credit access, (2) a univariate model with average household lending capacity ($HLC$) and (3) a benchmark model plus $HLC$. Each of the models has house prices ($HP$) as the dependent variable. The $\beta$ coefficients are optimised to minimise the squared prediction error (see Section \ref{sec:tuning}). 

\subsubsection{Benchmark Model}
The benchmark model is specified as follows: 

$$ HP_{t} = \beta_{0} + \beta_{1} I_{t} + \beta_{2} r_{t} + \beta_{3} LTV_{t} $$

\noindent where $I$ is quarterly income, $r$ are annual interest rates and $LTV$ is the average 'loan-to-value' ratio of first time buyers in quarter $t$.

\subsubsection{$HLC$ Model}
\begin{figure}[h]
\centering
  \includegraphics[width=0.7\linewidth]{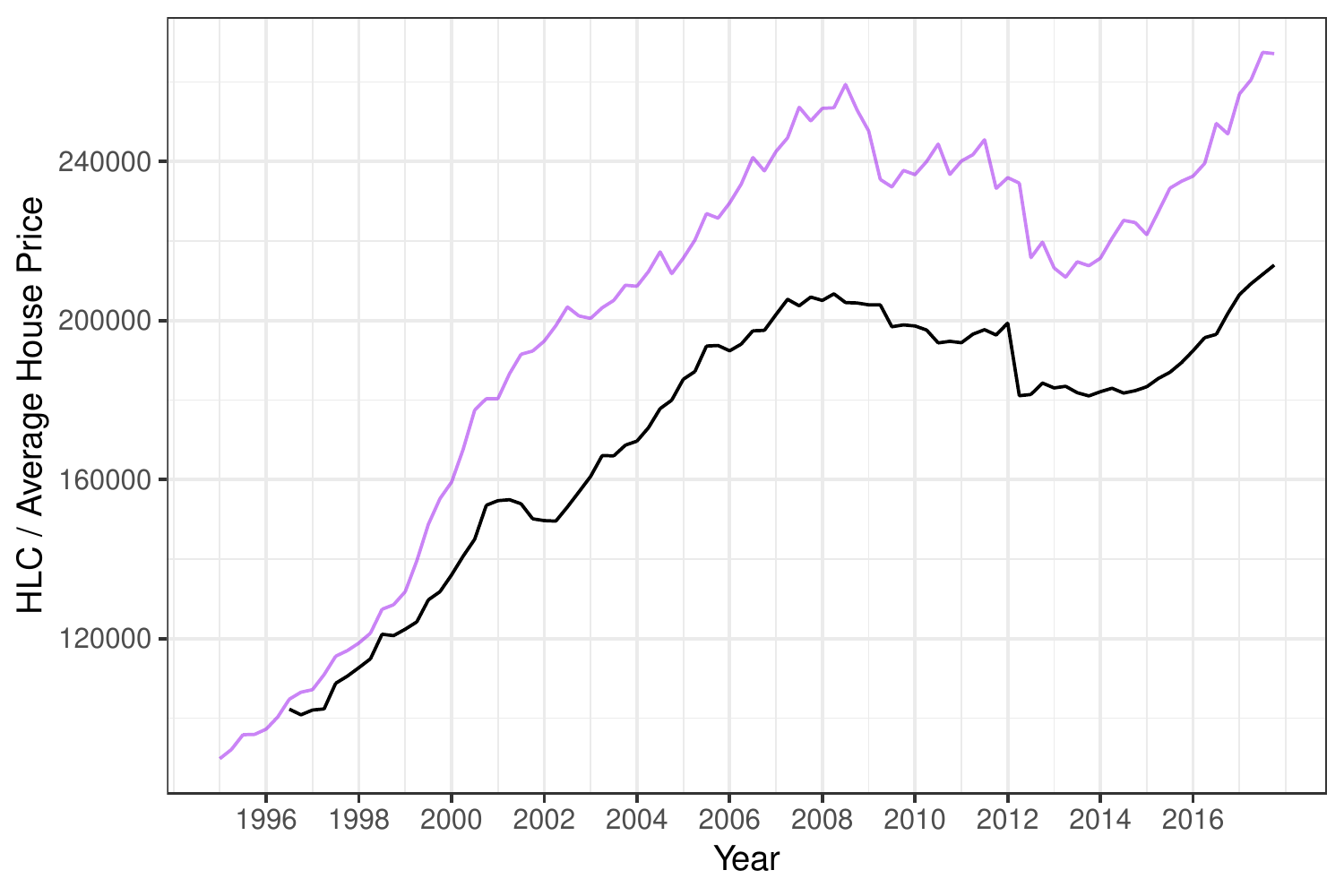}
  \caption{ The correlation between the average household lending capacity (black) and average house prices (purple) is strong. \label{fig:top_mortgage}}
\end{figure}

The univariate model only has $HLC$ as a lagged independent variable. $HLC$ is the amount banks allowed the average household to borrow at time $t$:

$$ HP_{t} = \beta_{0} + \beta_{1} HLC_{t-i} $$

\noindent where $HLC$ is constructed as a weighted average between the maximum interest-only mortgage ($HLC_{k}$) and the maximum annuity mortgage ($HLC_{a}$):

$$ HLC = m HLC_{k} + (1-m) HLC_{a} $$
$$ HLC_{k} = \frac{4 I W}{ (1 - x) r + c } $$
$$ HLC_{a} = \frac{I}{3} * W * a( (1 - x) r + c ) $$

\noindent where $m$ represents the market share of interest-only mortgages (see Section \ref{sec:marketshare}), $I$ is quarterly income\footnote{$I$ is multiplied by 4 to obtain the yearly income or divided by 3 to obtain monthly income.}, $W$ is the share of income used for housing expenses (see Section \ref{sec:W}), $x$ is the tax rate at which households can deduct interest (see Section \ref{sec:deduction}), $r$ are interest rates and $c$ is annual costs of maintaining the house expressed as a percentage of the purchase value (see Section \ref{sec:maintanance}). The lagged, unfitted variable $HLC$ is shown in Figure \ref{fig:top_mortgage}. The similarities in variance between $HLC$ and house prices is evident.

\subsubsection{Benchmark Plus $HLC$ Model}
The final model includes both benchmark variables and $HLC$. As $HLC$ is calculated based on income and interest rates, it naturally correlates highly with those variables. To reduce the cross-correlation, we convert $HLC$ by dividing $HLC$ by income $I$. This creates a new variable that expresses the ratio between lending capacity and income. This variable reflects the multiple of their disposable income that the average household can borrow. The benchmark model including $HLC$ has the form:

$$ HP_{t} = \beta_{0} + \beta_{1} I_{t} + \beta_{2} r_{t} + \beta_{3} LTV_{t} + \beta_{4} \frac{HLC_{t-i}}{I_{t-i}} $$

\subsection{Fitting Approaches}
The model specifications above have the form of an Ordinary Least Squares regression. OLS is the pocket knife of econometric modelling. It fits a minimum number of parameters and is highly interpretable but does not correct for auto correlation. The standard model in the literature that corrects for autocorrelation is the Error Correction Model, a more complex regression model that distinguishes between longterm and short-term predictors \citep{anundsen2013self, Turk2016}. ECM has the following generic form:

$$ \Delta y = \beta_{0} + \beta_{1} \Delta x_{1,t} + ... + \beta_{i} \Delta x_{i,t} + \gamma(y_{t-1} - (\alpha_{1} x_{1,t-1} + ... + \alpha_{i} x_{i,t-1}))$$

\noindent Predicting $\Delta y$ instead of $y$ indicates that we predict the change in $y$ rather than its absolute value. ECM was developed for econometric time-series analysis of variables that have trends on both the short and long term. ECM combines three estimates: it simultaneously estimates $y_{t-1}$ based on long-term predictors, $\Delta y$ based on short-term predictors and $\gamma$ to weight the longterm and short-term estimates. The strength of this model is that it anticipates the existence of long and short term trends and adjusts for auto correlation. However, the fact that the ECM model is much less sparse and includes a past value of the dependent variable in the fitting process makes it less persuasive. As it is the standard test in the literature, we fit all models as both OLS and ECM.

\subsection{Evaluating Model Performance \label{sec:evaluation}} 
We measure the quality of fit of every model with the Root Mean Squared Error (RMSE) and the Mean Absolute Error (MAE). We evaluate the tendency of the model to overfit spurious correlations in the data by performing two sets of fits. We also conduct an `out of sample` test, where the models are fitted only on data up to the second quarter of 2008. We choose that cut-off because it is the quarter in which the Dutch economy went into recession. The purpose of the test is to ascertain whether models are able to anticipate the coming crisis based on data from another part of the business cycle.

\section{Results \label{sec:results}}
In this section, we report the goodness of fit of 12 variants of the three model specifications above. We fit every model specification as an OLS and as an ECM on the all quarters and only quarters up to 2009. Tables \ref{tab:predict_benchmark}, \ref{tab:predict_hlc} and \ref{tab:predict_benchmark_hlc} present the measured quality of fit of all variants. Figure \ref{fig:predict_hp} displays the fitted values of the OLS models. We focus on the OLS results because the ECM models failed to generalise across the board. A figure with all the fitted values from the ECM models can be found in Appendix E.

\begin{figure}[p]
\vspace{3mm}
\minipage{0.45\textwidth}
\centering Fit On All Quarters \\
\endminipage\hfill
\minipage{0.45\textwidth}
\centering Out of Sample Test \\
\endminipage\hfill
\noindent\rule{16cm}{0.6pt}
\vspace{3mm}
\centering Benchmark \\
\vspace{3mm}
\minipage{0.45\textwidth}
  \includegraphics[width=\linewidth]{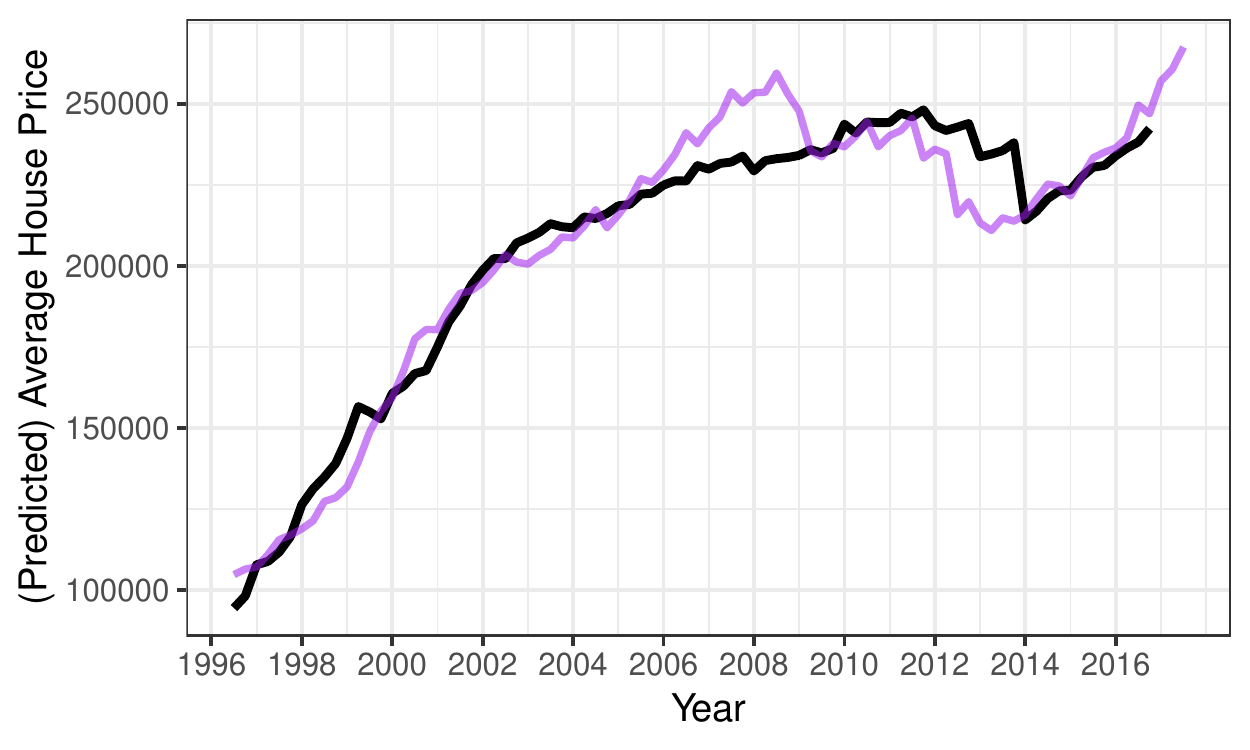}
\endminipage\hfill
\minipage{0.45\textwidth}
  \includegraphics[width=\linewidth]{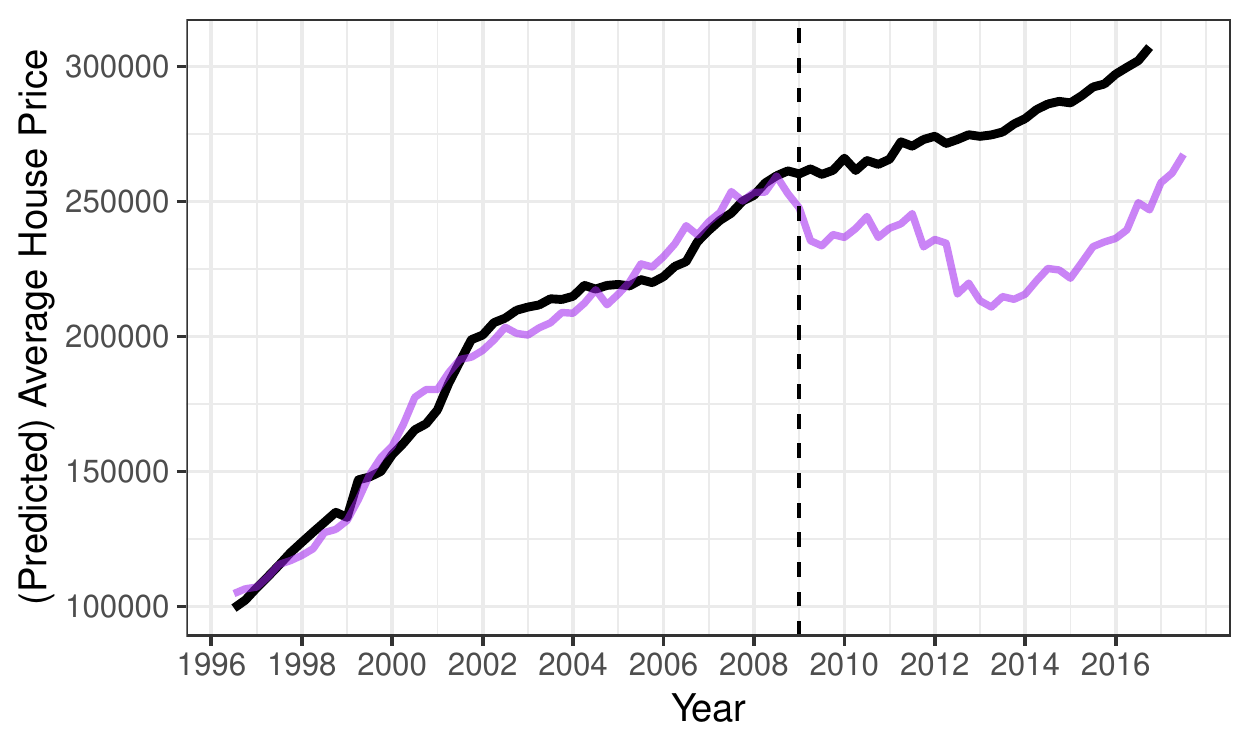}
\endminipage\hfill\newline
\noindent\rule{16cm}{0.6pt}
\vspace{3mm}
\centering $HLC$ \\
\vspace{3mm}
\minipage{0.45\textwidth}%
  \includegraphics[width=\linewidth]{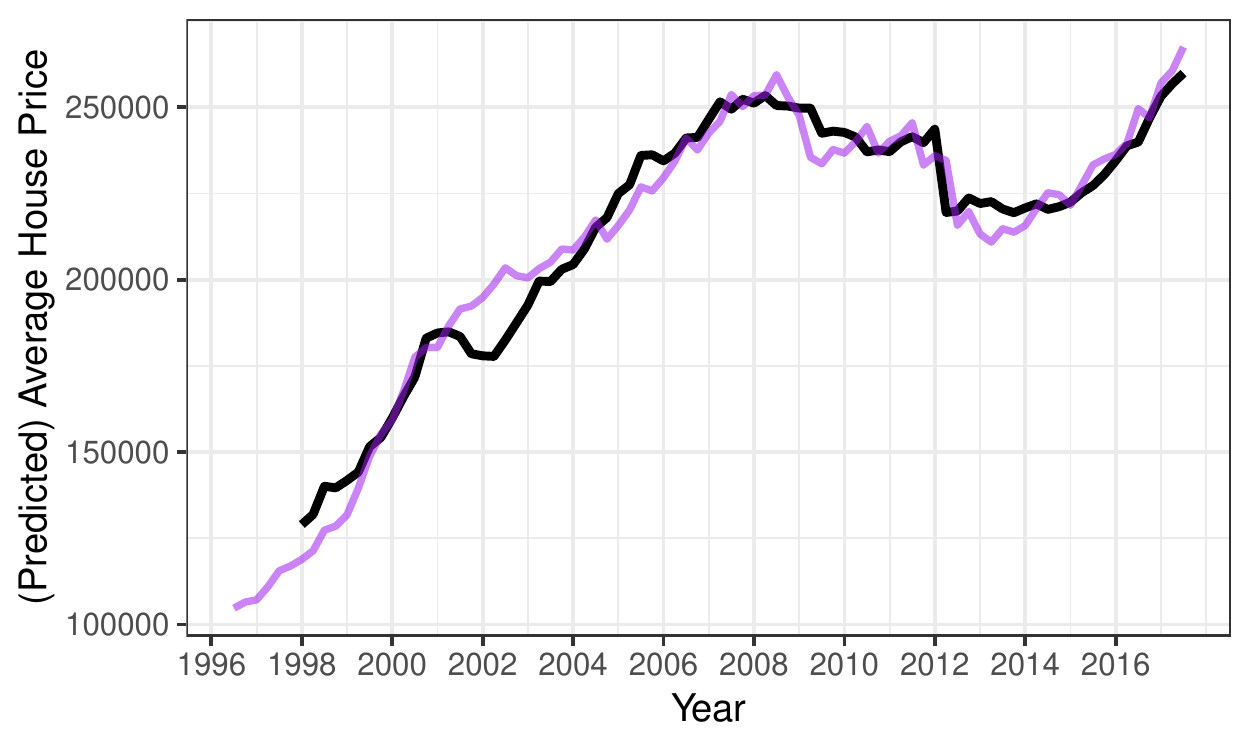}
\endminipage\hfill
\minipage{0.45\textwidth}
  \includegraphics[width=\linewidth]{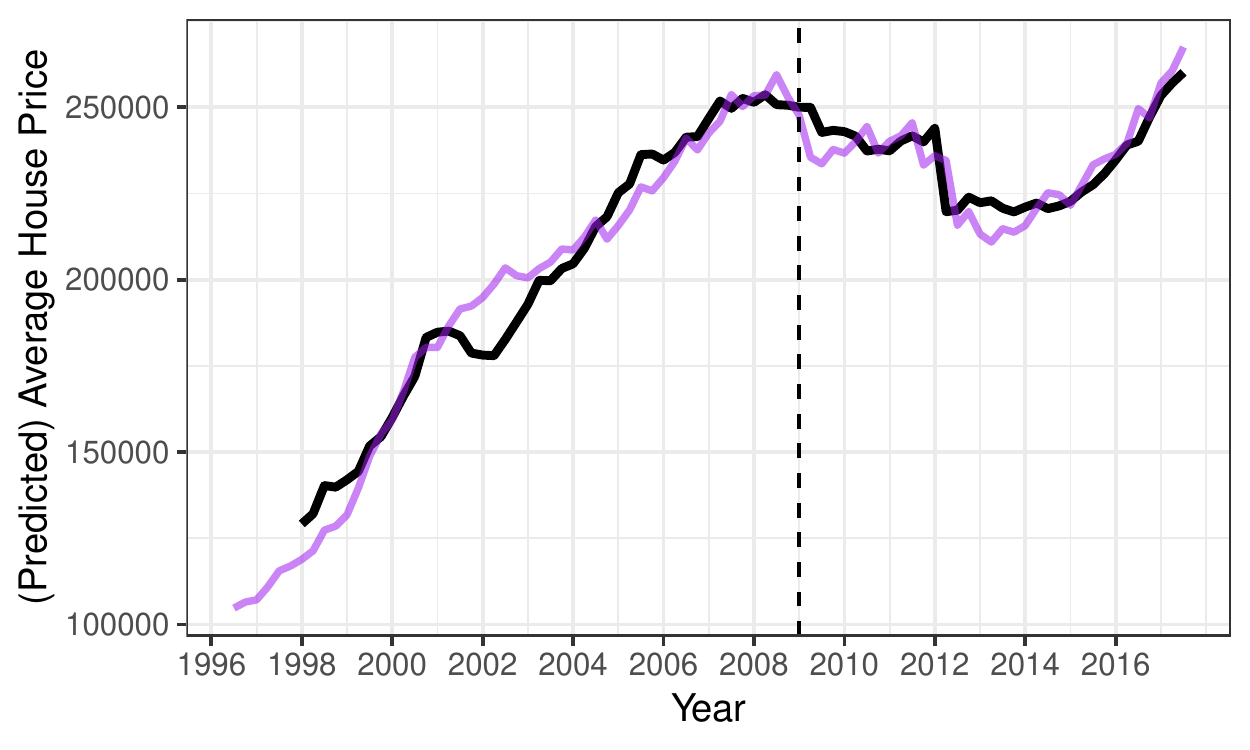}
\endminipage\hfill
\noindent\rule{16cm}{0.6pt}
\vspace{3mm}
\centering Benchmark incl. $HLC$ \\
\vspace{3mm}
\minipage{0.45\textwidth}%
  \includegraphics[width=\linewidth]{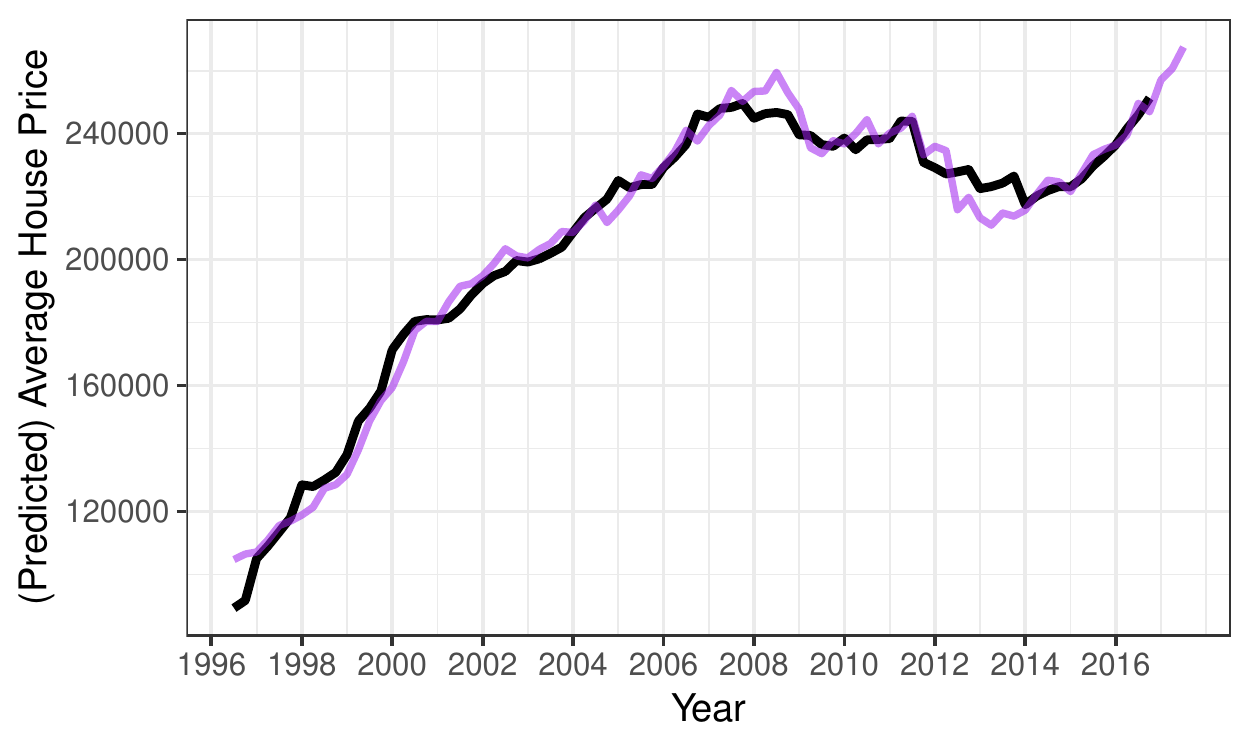}
\endminipage\hfill
\minipage{0.45\textwidth}
  \includegraphics[width=\linewidth]{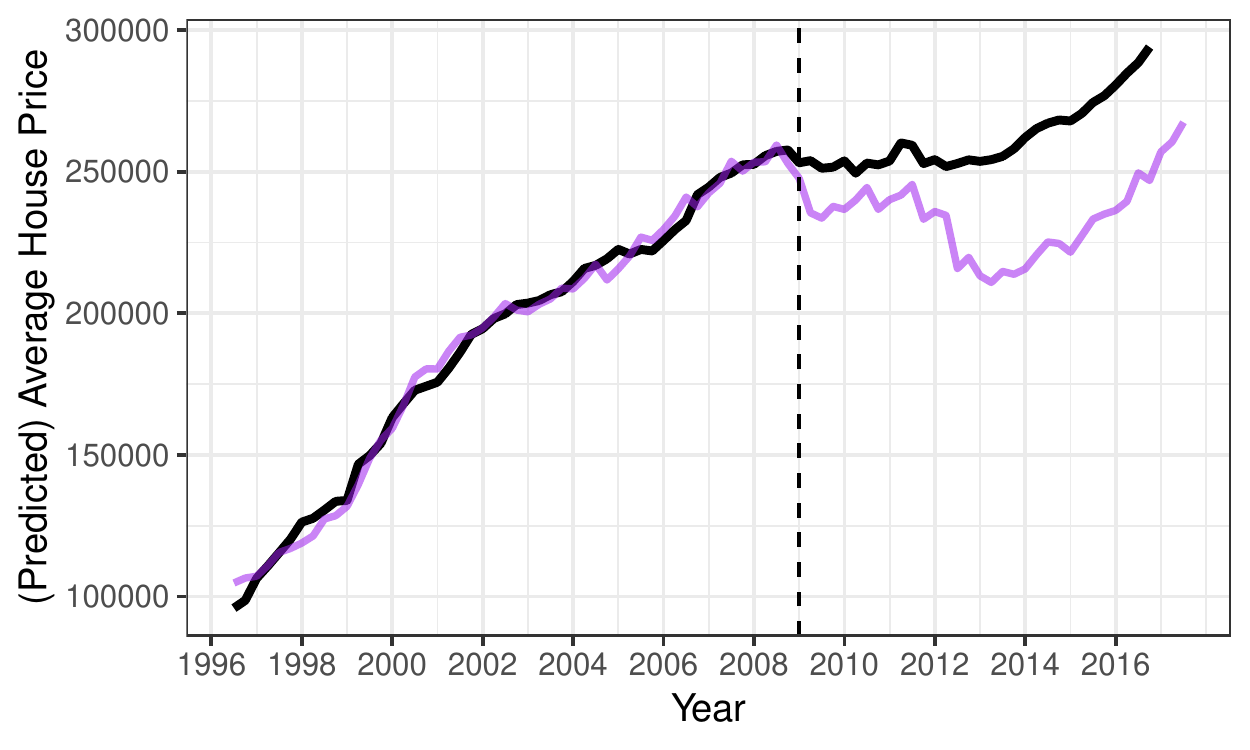}
\endminipage\hfill
\caption{Model Performance (OLS). These six diagrams displays the fitted values of regression models (black) contrasted with observed average house prices (purple). Fitted values on the left were generated by models fit on all the available data whereas fitted values on the right were generated by models trained only on quarters up to mid-2008. While the all three model specifications seem reasonably accurate when fit on all data, the `out-of-sample` test reveals that only the $HLC$ model forecasts accurately on unseen data.}
\label{fig:predict_hp}
\end{figure}

\newpage

\begin{table}[h]
\centering
  \caption{Benchmark Model}
  \label{tab:predict_benchmark}
\begin{tabular}{lccc}
\toprule
Model Type & RMSE & MAE\\
\midrule
\textit{Model: OLS} \\
Fit on  Quarters up to 2018 & 10.150 & 5.109 \\
Fit on Quarters up to 2009 & 31.173 & 8.146\\
\midrule
\textit{Model: ECM} \\
Fit on  Quarters up to 2018 & 8.057 & 5.152 \\
Fit on Quarters up to 2009 & 13.864 & 3.868 \\
\bottomrule
\end{tabular}
\end{table}

\begin{table}[h]
\centering
  \caption{Household Lending Capacity ($HLC$) Model}
  \label{tab:predict_hlc}
\begin{tabular}{lccc}
\toprule
Model Type & RMSE & MAE\\
\midrule
\textit{Model: OLS} \\
Fit on  Quarters up to 2018 & 7.451 & 4.891 \\
Fit on Quarters up to 2009 & 7.495 & 4.854 \\
\midrule
\textit{Model: ECM} \\
Fit on  Quarters up to 2018 & 6.700 & 4.681 \\
Fit on Quarters up to 2009 & 13.161 & 7.745 \\
\bottomrule
\end{tabular}
\end{table}

\begin{table}[h]
\centering
  \caption{Benchmark Plus $HLC$ Model}
  \label{tab:predict_benchmark_hlc}
\begin{tabular}{lccc}
\toprule
Model Type & RMSE & MAE\\
\midrule
\textit{Model: OLS} \\
Fit on  Quarters up to 2018 & 5.375 & 3.606 \\
Fit on Quarters up to 2009 & 21.210 & 5.744 \\
\midrule
\textit{Model: ECM} \\
Fit on  Quarters up to 2018 & 5.284 & 3.989 \\
Fit on Quarters up to 2009 & 8.664 & 2.994 \\
\bottomrule
\end{tabular}
\end{table}

\newpage

The most striking finding is that all but one models show poor performance in the out-of-sample test. Only the OLS fit of the $HLC$ model achieves comparable performance on the out-of-sample quarters and the in-sample quarters. By contrast, the benchmark models completely miss the direction of market in the out-of-sample quarters.\footnote{As a robustness test, mortgage debt-to-GDP ratios were also included in the benchmark. This version of the benchmark scored more poorly than the benchmark with LTV values.} The increase in robustness does not come at the expense of accuracy. When fit on all quarters, the OLS $HLC$ model has a 27\% lower MAE than the OLS benchmark model. The models that include both benchmark variables and $HLC$ have the best fit when fit on all quarters, but suffer from the same lack of robustness as the benchmark model.

\section{Discussion \label{sec:discussion}}
Economists agree that access to credit is a crucial determinant of house prices. This paper suggests a new approach to model credit conditions by modelling the structure of Loan-to-Income norms in the Netherlands. By modelling the formulas that Dutch banks use to calculate how much they can lend to households, we construct a regression model that does not overfit in the `out-of-sample` test. We compare the accuracy of the model with existing measures of credit access and find it delivers more accurate results. Our analysis indicates that the cause of the Dutch crisis was a Dutch phenomenon: the rise of interest-only mortgages and their fall at the hands of the Dutch government in 2011. As the rise and fall of these mortgage products is correlated with rising and falling LTV rates, this effect is partly represented in the benchmark model, but modelling it explicitly allows for a better fit. The findings suggest that the `double dip` crisis was partly due to the fact that Dutch government constricted lending capacity in 2011 by regulating interest-only mortgages more strictly. Whereas all economic indicators pointed towards price recovery, prices fell steeply soon after this legislation came into effect.

More significantly, the model specification based on LTI formulas appears to generalise more robustly when only trained on a fraction of the available data. This observation suggests that the in-sample accuracy of econometric house price models  may be a poor measure of their predictive power. These models may be fitting spurious correlations that do not generalise to other economic circumstances. The field faces a challenge: how do we develop models that account for complex interactions between economic variables (low bias) whilst safeguarding against overfitting (low variance)? The methodology of this work lays out a possible path forward. We identify relationships between variables in qualitative research and construct a measure based on those relationships for quantitive analysis. Rather than fit each variable independently\footnote{In the general form $ y = \beta_{0} + \beta_{1}x_{1} + ... + \beta_{i}x_{i} $.}, we fit a model where a function based on the independent variables is fitted\footnote{In the general form: $ y = \beta_{0} + \beta_{1}f(x_{1}, ... , x_{i}) $}. In this work, the function calculates household lending capacity based on the formulas that Dutch banks use in the mortgage application process. Others authors can represent any hypothesised interaction between variables as functions. Thus, one can model any complex relationship between variables, hence lowering bias, whilst fitting fewer parameters, hence lowering variance. This route might offer an alternative route to the current direction of the field: plugging an increasing number of variables into increasingly complex statistical models.\footnote{A possible critique of this method would be that it invites researchers to formulate endlessly complex versions of $f$ until they find some function that fits the dependent variable. To counter this, researchers should be required to carefully justify their hypothesised function based on qualitative research. In this work, we base the construction of $HLC$ on the LTI formulas of Dutch banks. Both the theoretical justification of the construction of $f$ \emph{and} the empirical validation must be persuasive.}

\subsection*{Competing interests}
The authors declare that they have no competing interests.

\subsection*{Funding}
TY was partially supported by the Alan Turing Institute under the EPSRC grant no. EP/N510129/1. The sponsor had no role in study design; in the collection, analysis and interpretation of data; in the writing of the report; and in the decision to submit the article for publication.

\subsection*{Authors' contributions}
MS analyzed the data. MS and TY designed the study and the analysis. All authors read and approved the final manuscript.

\bibliographystyle{jss2}
\bibliography{thesis.bib}

\newpage

%\doublespacing
\appendix

\section{Summary of Dutch Housing Market History}

This section summarises the Dutch story, highlights existing theories and identifies the gap in understanding.
The Dutch have consistently seen house prices rise in the post-war period. There has a been a near-permanent shortage of residential properties, so supply rarely meets demand.\footnote{This is a reason that supply-side factors are less relevant in the Dutch market.} In a country with one of the highest population densities in the world, population growth and rising incomes, households expect that residential property values will continue to rise. This makes property a good investment - especially in popular cities such as Amsterdam, Utrecht and The Hague. 

Figure \ref{fig:NedStats} shows that house prices rose rapidly in the 1990's and early 2000's and entered a period of sustained decline after 2009. Despite falling interest rates, housing prices fell 16\% from their peak value. The crisis came at great cost to many Dutch families. In 2015, 28\% of homes were deemed `under water': the home was worth less than the outstanding mortgage debt \citep{CPB2014}. Families could not move to a new house because they could not pay off the debt on their old property. Divorcees, the recently unemployed and other citizens with an urgent reason to sell their house could not fetch a price higher or equal to their mortgage.  As a result, these citizens were left with substantial and long-term debts without the means to repay them. Often, the costs of new housing and debt payments were more than these households could afford.

The Dutch government attempted to boost home prices by lowering taxes on real-estate transactions and allowing parents to assist their children to buy property by allowing a tax-free gift for the purpose of acquiring a property. Interest rates fell to record lows \citep{DNBr}. However, the housing market continued to slump. After years of falling prices, the market rose in 2015. Prices shot back up beyond their previous peak within a few years. In the large cities, prices rose with 10\% per year, leading the Dutch Central Bank to call markets ``overheated'' that had been `cold' just a few years before. The Dutch Central Bank expressed concern about the unprecedented speed of price increases \citep{Groot2018}. 

The question remains which factors drove this dramatic cyclical movement in the market. Figure \ref{fig:NedStats} shows that The Netherlands experienced a `double dip' in prices: prices fell in 2008 and 2009, stabilized from 2009 to 2011 and sank further in 2013. The second dip occurred despite the fact that Gross Domestic Product had resumed to grow (see Figure \ref{fig:NedStats}). Research remains to be done why the Dutch market experienced a `double dip' that prolonged the crisis on the housing market. 

The International Monetary Fund, the Dutch National Bank and the Centraal Plan Bureau have performed a great number of analyses of the housing market. These institutions express concern that Dutch households are amongst the most indebted in the world \citep{IMF2019Netherlands}. The bulk of household debt consists of mortgage debts. Dutch households borrow because of fiscal incentives to do so. Households either rent or own property. Buying property with a mortgage is attractive, because monthly payments are partially a form of savings, whereas rent is not. The Netherlands makes it fiscally attractive for households to switch from renting to owning property. This has three causes. The first is that renters in the middle and upper class pay some of the highest rents in Europe\footnote{The cause of high rents in the Netherlands is a topic of much debate and beyond the scope of this work.}. The second is that the government allows home-owners to deduct interest from their taxable income, effectively subsidising mortgage debt. The third cause is that the Dutch government guarantees mortgage debt for homes valued up to \euro290.000 under the `Nationale Hypotheek Garantie' (NGH) scheme. This reduces risks for banks and thus allows them to provide lower interest rates on middle class mortgages. NHG allows home owners to default on mortgage debt that cannot be paid off by selling their own, reducing risks for new home owners. In summary, households realise that the most fiscally profitable route is to buy property using the highest amount of mortgage debt they can obtain.

\newpage

\section{Data Sources}

\subsection{House Prices}
We collect the average transaction price of residential properties in the Netherlands ($HP$) as reported on a quarterly basis by the Dutch statistics office \citep{CBSHP}. Values before 1999 were only reported on a yearly basis. For these years, we imputed the missing quarters by extrapolating the trend between years. While this approach may miss quarterly variation, it captures the overall trend in prices.

\subsection{Household Income}
Household Income after Taxes ($I$) is average after-tax household income as reported by the Dutch statistics office \citep{CBSI}. The variable has strong yearly seasonality, because bonuses are added to income in the last quarter. Therefore, we smooth the variable by taking the average of the last 4 quarters. As a result, changes in the variable show changes in year-over-year income.

\subsection{Interest Rates}
Banks calculate lending capacity using the nominal interest rate for the mortgage product. Mortgage interest rates differ from the interest rates on the capital markets, because banks factor in their margin, estimated risk of default, risk of devaluation of the property and future developments of the interest rates. Every bank sets different interest rates for their products. We obtained the average mortgage interest rate for new contracts on a quarterly basis from the Dutch Central Bank for the period 1992-2019 \citep{DNBr}. We smooth the variable by taking the average of the last 4 quarters. Changes in the variable show changes year-over-year.

\subsection{Average LTV Values} Average LTV ratios were collected from reports by the Dutch Central Bank and the Rabobank \citep{DNB2019LTV, Rabo2015}. The data was reported on a yearly level, hence we miss some quarter-to-quarter variance. Where data was missing, we imputed the value from the year before.

\subsection{Market Share of Interest Only Mortgages} Market Share of Interest-Only mortgages was collected from two sources: a report by the Dutch Central Bank up to 2008 \citep{DNBm} and a Rabobank report up to 2011 \citep{RABOm} after which the category became irrelevant.

\newpage

\section{Setting Constants in Model Construction}

\subsection{Accounting for Mortgage Interest Deduction ($t$) \label{sec:deduction}} The Dutch government allows to deduct mortgage payments from income tax. Thus, the effective interest payed is a fraction of the nominal interest rate. For example, a household might pay a top rate of 50\% tax over their income. If they have \euro10.000 in mortgage interest payments, they can subtract this sum from their taxable income. They do not have to pay 50\% tax over \euro10.000 of their total income and as a result `save' \euro5.000. This example demonstrates that the amount of mortgage interest deduction depends on the top rate that households pay. The Dutch income tax system features many tax brackets and deductions. The top marginal rate can differ between households with the same income. As tax policy has been adjusted, the brackets have also shifted over time. The minutiae of Dutch income tax policy were beyond the scope of this work to model and we take a simplified approach. The average household has traditionally paid their top rate in a tax bracket of approximately 40\% . We assume that households deduct their interest payments in the 40\% bracket. The household pays 60\% of their interest payment; the rest is subsidised by the state. Thus, $t$ is set at 0.4.

\subsection{Share of Income for Housing Expenses ($W$) \label{sec:W}} The `woonquote' governs the share of income the bank assumes that buyers can use to pay for the costs associated with their home. The precise `woonquote' used for the average Dutch household is not known, because the rules and regulations only defines a range of 21\% to 40\% \citep{NIBUD2018}. Banks can choose a value within that range based on a large number of variables not available to me, such as family size, projected (energy) costs of the home, projections of future income, etc. It is possible that the `woonquote' was not the same for interest-only mortgages and annuities, because these mortgages may have been chosen by differing groups whose characteristics put them elsewhere on the range. Without any hard data, we choose to set the parameter in the middle of the range: 30\%. If the coefficient is not set appropriately, this will be compensated for in the estimation of $\beta_{1}$.

\subsection{Setting Maintenance Cost ($c$) \label{sec:maintanance}} LTI limits specify that homeowners must be able to pay all their costs from the portion of their income allocated for housing. We set the `other costs` parameter at 2,5\% of the value of the home. The NIBUD - the government agency that sets LTI rules - estimates `other costs' to be a significant fraction of the initial cost of the home. This matches a back-of-the-envelope calculation: homeowners face a tax of 0,75\% of their home value, municipal taxes and approximately 1\% yearly costs for maintanance. 

\subsection{Constructing Marketshare ($m$) \label{sec:marketshare}} For the construction of $HLC$, we require the market share of interest-only mortgage over time. That data was not available over the entire time period. However, we estimate the market share of interest-only mortgages in new transactions based on the changes in the total stock of Dutch mortgages. The rest of this section lays out how this measure was constructed.

We can approximate the share of households who got a new mortgage by dividing the number of market transactions by the total number of households for every year. This assumes that no household moved twice. Based on the overall marketshare data, we calculate the increase in market share for interest-only mortgages. If we know that 5\% of households moved in a given year and interest-only mortgages captured 2,5\% of the market, we can deduce that 50\% of households who moved switched from an annuity-like mortgage to an interest-only mortgage. What about the remaining 50\%? These must be non-switchers. We can assume that non-switcher households renewed the mortgage type they already possessed. Therefore, in a case where interest-only mortgages have a marketshare of 40\%, we can deduce that the share of non-switcher households who renewed an interest-only mortgage was the proportion of non-switchers times the overall market share ($0.5 *0.4 = 0.2$). In sum, we can deduce that 50\% of households switched into interest-only, 20\% of households renewed an interest-only mortgage and 30\% renewed an an annuity-like mortgage. Thus, the market share of interest-only mortgages of new mortgages is 70\%.

\subsection{Calculating Optimal Lag for $HLC$ \label{sec:timelag}}
This section concerns itself with choosing an appropriate time-lag for $HLC$ in our regression analysis. In time-series analysis, the causal relationship between variables might be difficult to observe because of a time-lag. In those cases, we can shift the independent variables backward in time. If we believe that lending capacity today will influence housing prices 2 years from now, we regress lending capacity today against housing prices in two years. Timelags - like most other causal relationships -  can be chosen on purely theoretical grounds or inferred from the data. we employ a common, hybrid approach. We first specify a range of possible lags based on theory. This assures that any of the lags would accord with the theoretical conception of the behaviours of home-buyers. We then test empirically which lag best fits the data. We regress the different lagged versions of $HLC$ against house prices and take the $R^{2}$ to measure goodness of fit. We choose the lag with the highest $R^{2}$. 

\begin{figure}[h]
\centering
  \includegraphics[width=0.6\linewidth]{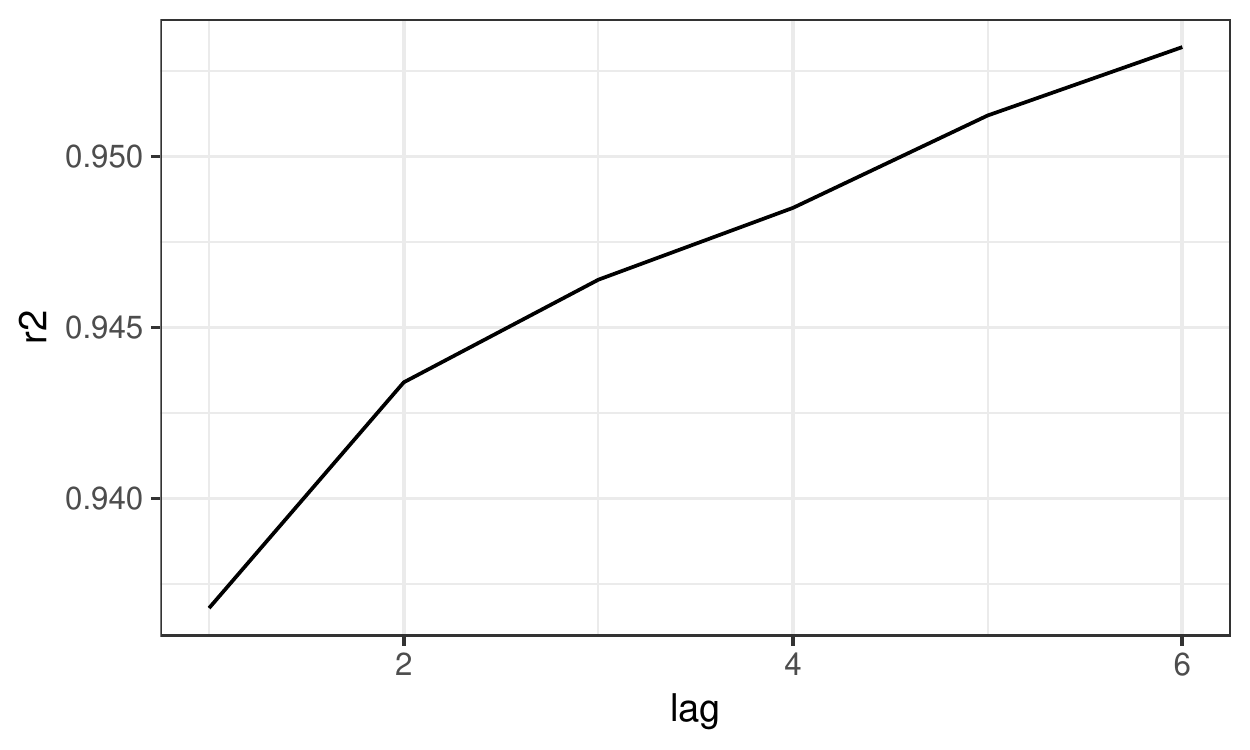}
  \caption{ $R^{2}$ of $HLC$ with different lags.}
\end{figure}

We chose a range of 0-6 quarters to lag $HLC$ by, because homeowners often arrange financing first and then acquire a home. The process of acquiring a home often involves bidding, bargaining, a waiting period for the owners to find a new home and finally the formal transaction.  That process can take many months. Thus, we choose the range of 0-6 quarters as possible lags between lending capacity and house prices. We regressed lending capacity lagged by 0-6 quarters against house prices and observed the $R^{2}$. The $R^{2}$ is highest with a time-lag of 6 quarters.\footnote{We also tested lags greater than 6 and found the $R^{2}$ is lower with a lag of 7 and 8 quarters.}

\subsubsection{Choosing the number of coefficients to estimate \label{sec:tuning}} Originally, we  intended to fit separate coefficients for $HLC_{a}$ and $HLC_{k}$. However, we found that - remarkably - the value of both coefficients was nearly identical (on three decimals) for the optimal fit. This suggests that the best fit parameter values of constants $c$, $t$ and $W$ that are identical for $HLC_{a}$ and $HLC_{k}$. For the sake of representing our model as clearly as possible, we therefore chose to simplify the model and only formally estimate one coefficient on $HLC$ in the regression analysis. However, for other values of $c$, $t$ and $W$ or in other markets, it may be necessary to fit separate $\beta$'s for different mortgage product types.

\newpage

\section{Model Fits}

% Table created by stargazer v.5.2.2 by Marek Hlavac, Harvard University. E-mail: hlavac at fas.harvard.edu
% Date and time: Sat, Jul 20, 2019 - 21:45:05
\begin{table}[!htbp] \centering 
  \caption{OLS Results} 
  \label{tab:ols_reg} 
\begin{tabular}{@{\extracolsep{5pt}}lccc} 
\\[-1.8ex]\hline 
\hline \\[-1.8ex] 
 & \multicolumn{3}{c}{\textit{Dependent variable:}} \\ 
\cline{2-4} 
\\[-1.8ex] & \multicolumn{3}{c}{$HP$} \\ 
\\[-1.8ex] & (1) & (2) & (3)\\ 
\hline \\[-1.8ex] 
$I$ & 3.618$^{***}$ &  & 4.145$^{***}$ \\ 
  & (0.218) &  & (0.073) \\ 
  & & & \\ 
$LTV$ & 6,070.454$^{***}$ &  & 2,604.368$^{***}$ \\ 
  & (519.257) &  & (326.972) \\ 
  & & & \\ 
$r$ & $-$6,270.771$^{***}$ &  &  \\ 
  & (2,167.802) &  &  \\ 
  & & & \\ 
$HLC_{t-6}$&  &  1.410$^{***}$ &  \\ 
  &  & (0.058) &  \\ 
  & & & \\ 
 $\frac{HLC}{I}_{t-6}$ &  &  & 67,784.850$^{***}$ \\ 
  &  &  & (4,132.603) \\ 
  & & & \\ 
 Constant & $-$627,143.700$^{***}$ & $-$39,257.250$^{***}$ & $-$521,212.700$^{***}$ \\ 
  & (53,894.820) & (10,537.810) & (30,881.110) \\ 
  & & & \\ 
\hline \\[-1.8ex] 
Observations & 82 & 81 & 78 \\ 
R$^{2}$ & 0.937 & 0.953 & 0.980 \\ 
Adjusted R$^{2}$ & 0.934 & 0.952 & 0.979 \\ 
Residual Std. Error & 11,056.030 (df = 78) & 8,392.266 (df = 79) & 5,519.168 (df = 74) \\ 
F Statistic & 385.834$^{***}$ (df = 3; 78) & 1,590.978$^{***}$ (df = 1; 79) \ & 1,180.502$^{***}$ (df = 3; 74) \\ 
\hline 
\hline \\[-1.8ex] 
\textit{Note:}  & \multicolumn{3}{r}{$^{*}$p$<$0.1; $^{**}$p$<$0.05; $^{***}$p$<$0.01} \\ 
\end{tabular} 
\end{table} 

% Table created by stargazer v.5.2.2 by Marek Hlavac, Harvard University. E-mail: hlavac at fas.harvard.edu
% Date and time: Sat, Jul 20, 2019 - 21:36:15
\begin{table}[!htbp] \centering 
  \caption{ECM Results} 
  \label{tab:ecm_reg} 
\begin{tabular}{@{\extracolsep{5pt}}lccc} 
\\[-1.8ex]\hline 
\hline \\[-1.8ex] 
 & \multicolumn{3}{c}{\textit{Dependent variable:}} \\ 
\cline{2-4} 
\\[-1.8ex] & \multicolumn{3}{c}{$\Delta HP$} \\ 
\\[-1.8ex] & (1) & (2) & (3)\\ 
\hline \\[-1.8ex] 
$\Delta I$ & 1.241 &  & 1.068 \\ 
  & (1.013) &  & (0.958) \\ 
  & & & \\ 
$\Delta LTV$ & $-$595.012 &  & $-$77.518 \\ 
  & (2,244.856) &  & (2,085.333) \\ 
  & & & \\ 
 $\Delta r$ & 161.405 &  & $-$1,493.893 \\ 
  & (3,958.949) &  & (3,681.579) \\ 
  & & & \\ 
$\Delta \frac{HLC}{I}_{t-4}$ &  &  & 23,430.740$^{*}$ \\ 
  &  &  & (12,426.520) \\ 
  & & & \\ 
$I_{t-1}$ & $-$0.098 &  & 1.086$^{***}$ \\ 
  & (0.219) &  & (0.371) \\ 
  & & & \\ 
$LTV_{t-1}$ & 454.846 &  & 249.341 \\ 
  & (376.520) &  & (353.242) \\ 
  & & & \\ 
$r_{t-1}$ & $-$3,753.088$^{***}$ &  & $-$265.662 \\ 
  & (1,077.769) &  & (1,396.372) \\ 
  & & & \\ 
$\Delta HLC_{t-4}$ &  & 1.074$^{***}$ &  \\ 
  &  & (0.239) &  \\ 
  & & & \\ 
 $HLC_{t-5}$ &  & 0.259$^{***}$ &  \\ 
  &  & (0.092) &  \\ 
  & & & \\ 
  $\frac{HLC}{I}_{t-5}$ &  &  & 26,801.280$^{***}$ \\ 
  &  &  & (7,095.525) \\ 
  & & & \\ 
$\gamma$ & $-$0.090$^{*}$ & $-$0.197$^{***}$ & $-$0.314$^{***}$ \\ 
  & (0.053) & (0.065) & (0.080) \\ 
  & & & \\ 
 Constant & 468.108 & $-$3,439.291 & $-$102,568.800$^{**}$ \\ 
  & (38,566.940) & (3,583.361) & (44,634.110) \\ 
  & & & \\ 
\hline \\[-1.8ex] 
Observations & 83 & 83 & 83 \\ 
R$^{2}$ & 0.246 & 0.316 & 0.374 \\ 
Adjusted R$^{2}$ & 0.175 & 0.290 & 0.297 \\ 
Residual Std. Error & 4,453.263 (df = 75) & 4,130.705 (df = 79) & 4,112.399 (df = 73) \\ 
F Statistic & 3.490$^{***}$ (df = 7; 75) & 12.188$^{***}$ (df = 3; 79) & 4.844$^{***}$ (df = 9; 73) \\ 
\hline 
\hline \\[-1.8ex] 
\textit{Note:}  & \multicolumn{3}{r}{$^{*}$p$<$0.1; $^{**}$p$<$0.05; $^{***}$p$<$0.01} \\ 
\end{tabular} 
\end{table}

\begin{figure}[p]
\minipage{0.4\textwidth}
\centering Fit On All Quarters \\
\endminipage\hfill
\minipage{0.4\textwidth}
\centering Out of Sample Test \\
\endminipage\hfill
\noindent\rule{16cm}{0.6pt}
\vspace{3mm}
\centering Benchmark \\
\vspace{3mm}
\minipage{0.4\textwidth}
  \includegraphics[width=\linewidth]{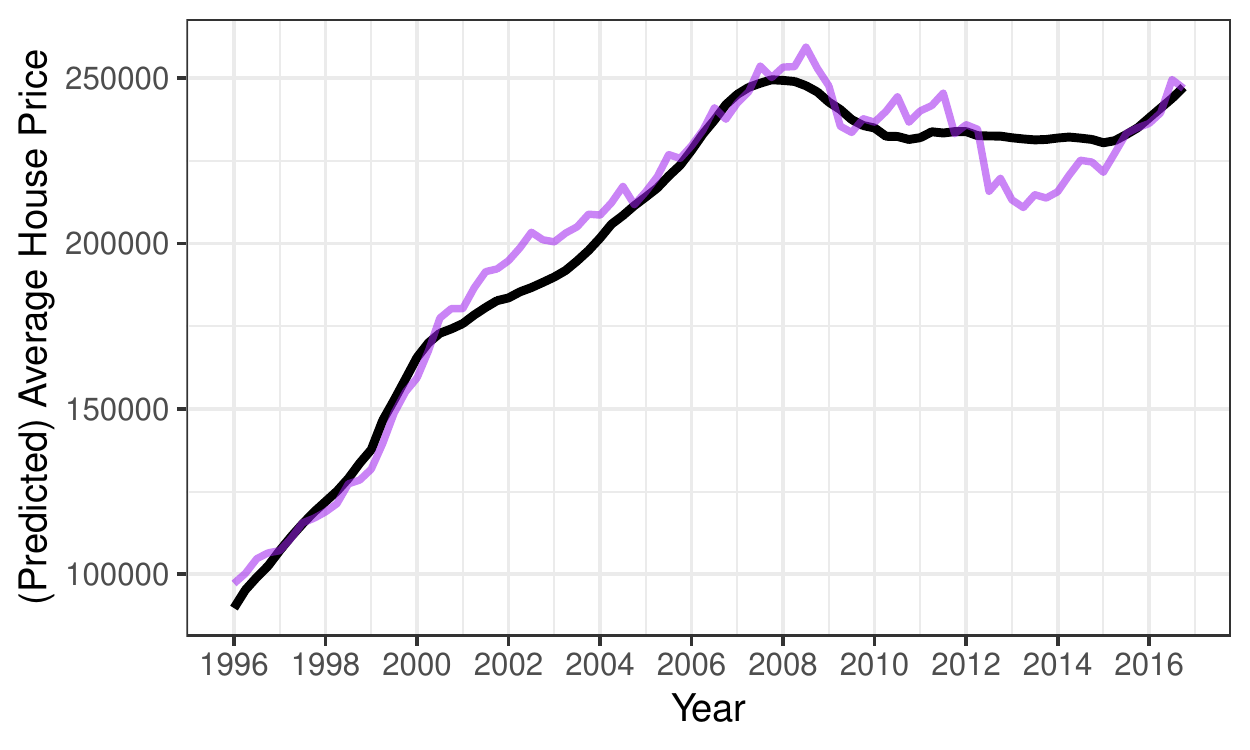}
\endminipage\hfill
\minipage{0.4\textwidth}
  \includegraphics[width=\linewidth]{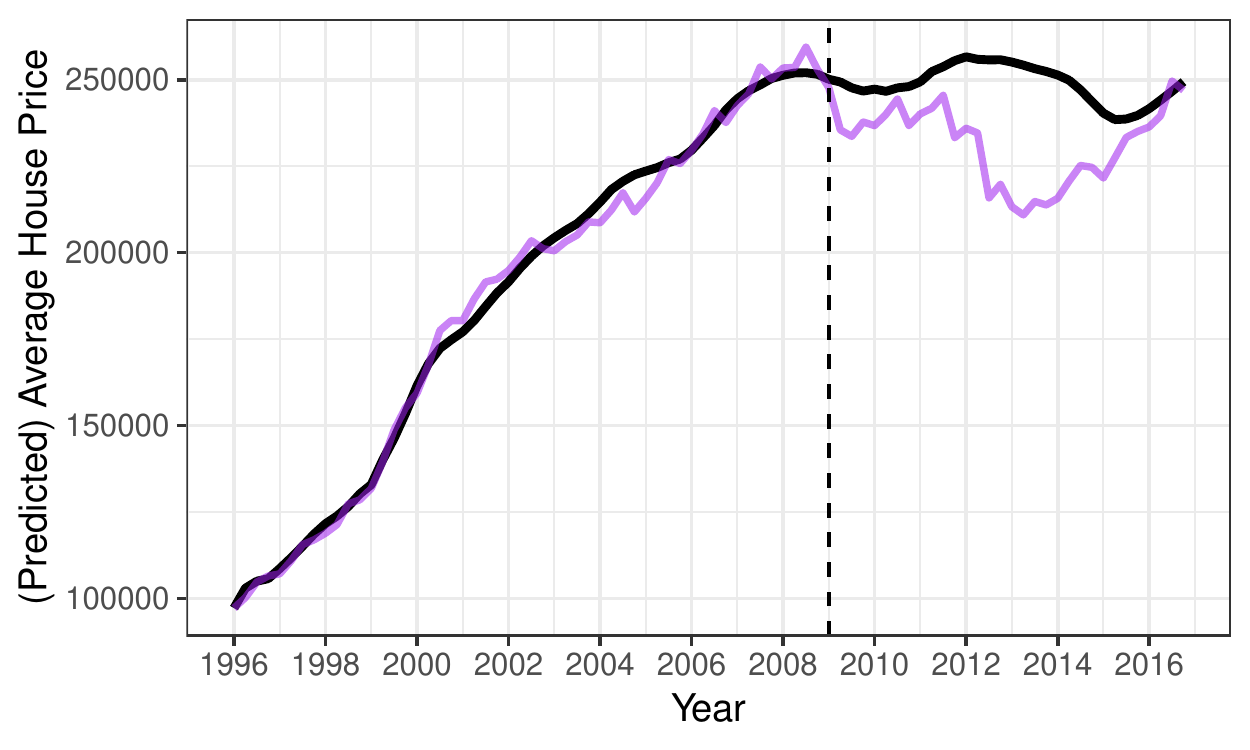}
\endminipage\hfill\newline
\noindent\rule{16cm}{0.6pt}
\vspace{3mm}
\centering $HLC$ \\
\vspace{3mm}
\minipage{0.4\textwidth}%
  \includegraphics[width=\linewidth]{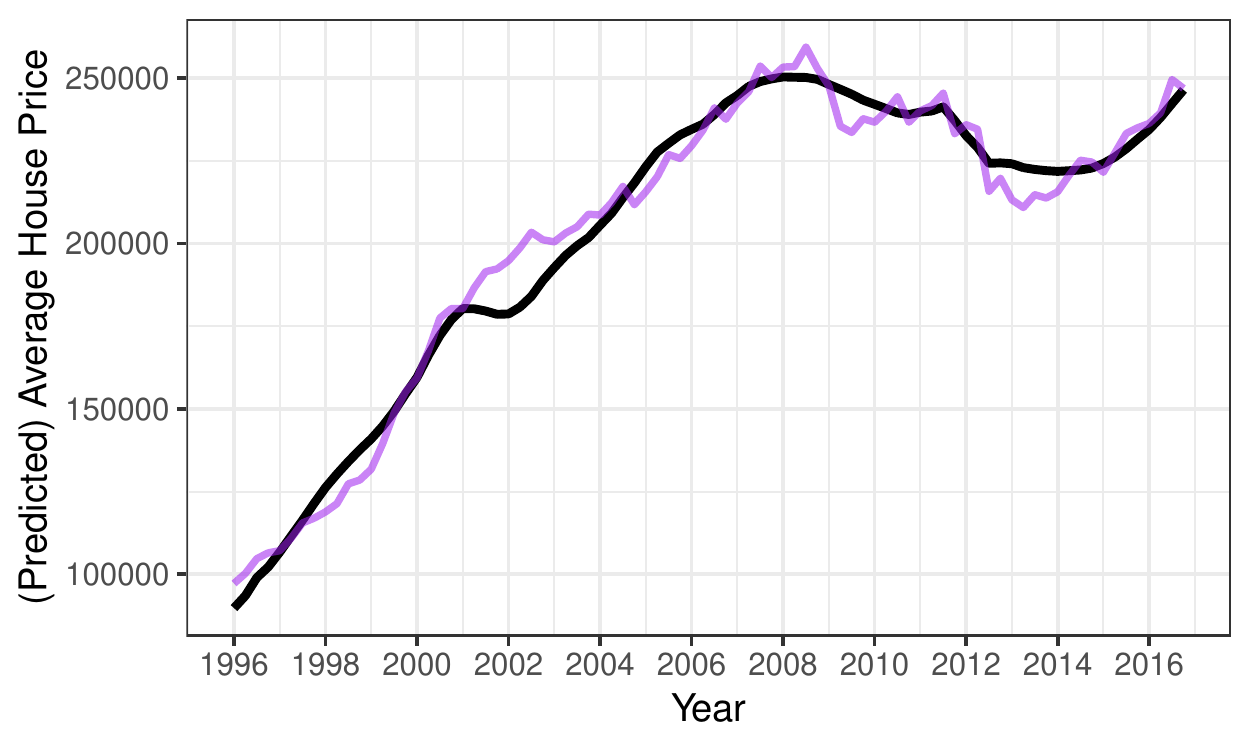}
\endminipage\hfill
\minipage{0.4\textwidth}
  \includegraphics[width=\linewidth]{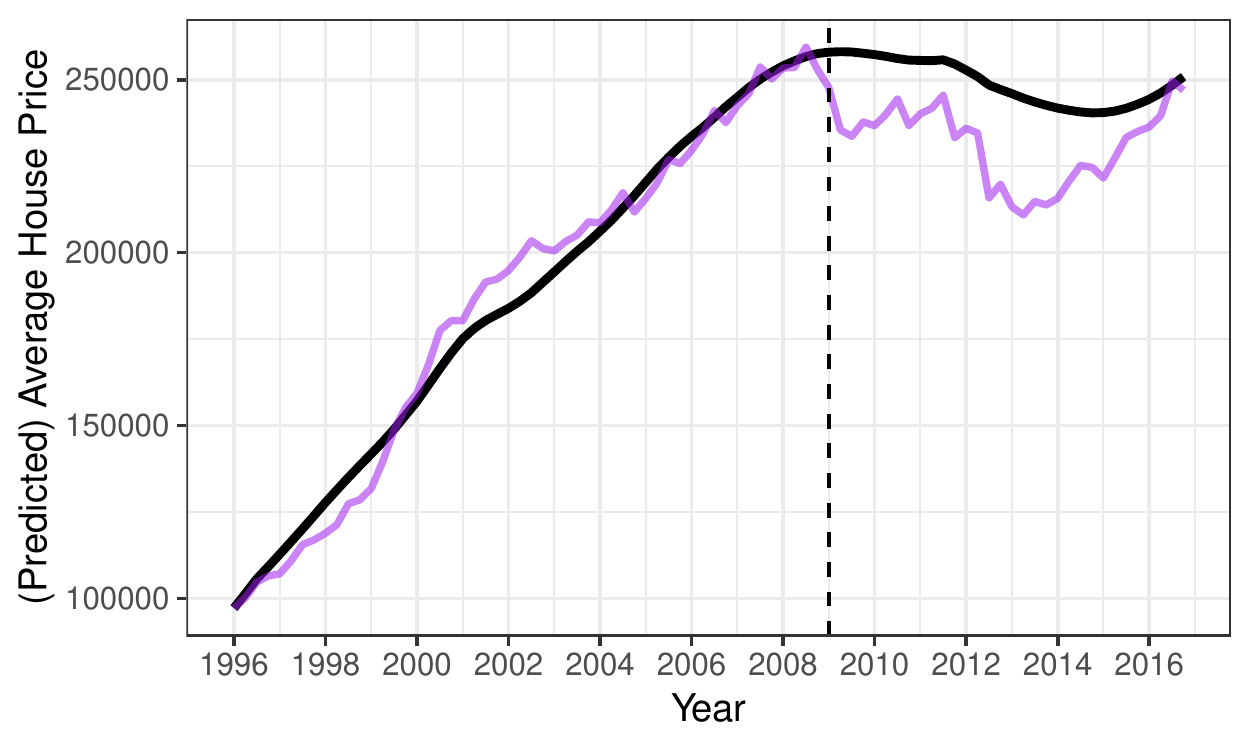}
\endminipage\hfill
\noindent\rule{16cm}{0.6pt}
\vspace{3mm}
\centering Benchmark incl. $HLC$ \\
\vspace{3mm}
\minipage{0.4\textwidth}%
  \includegraphics[width=\linewidth]{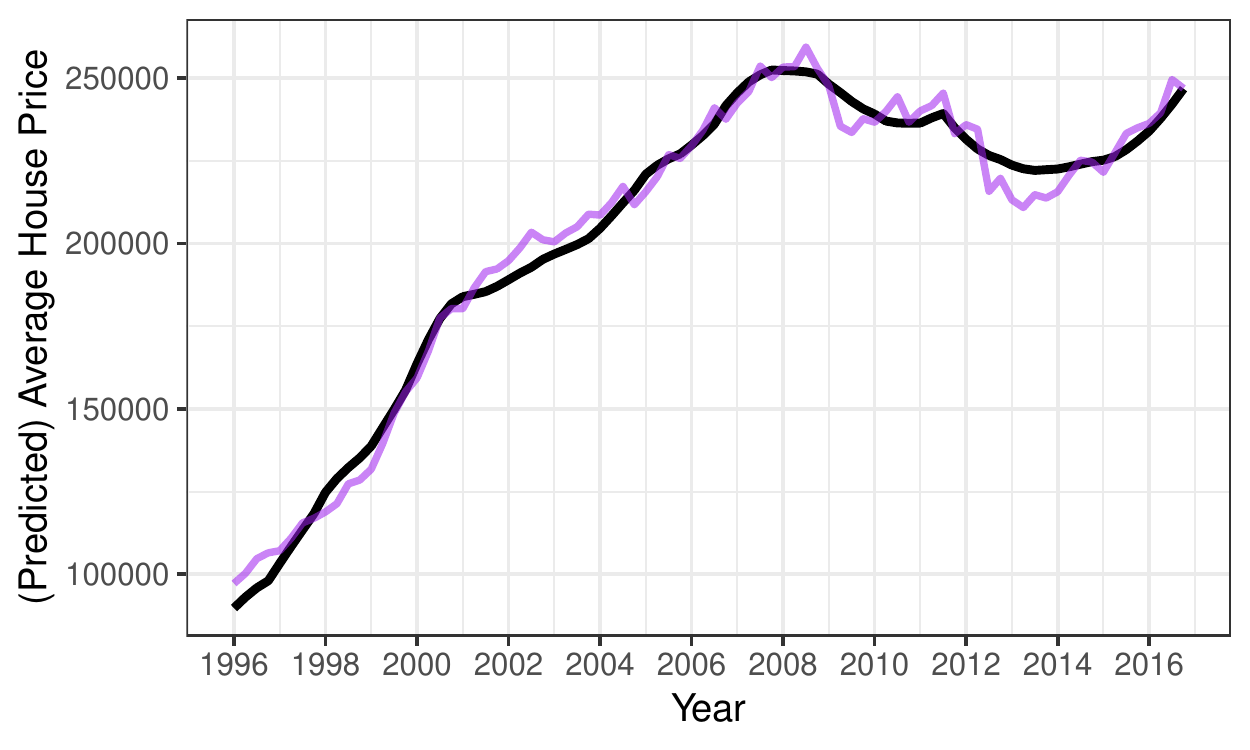}
\endminipage\hfill
\minipage{0.4\textwidth}
  \includegraphics[width=\linewidth]{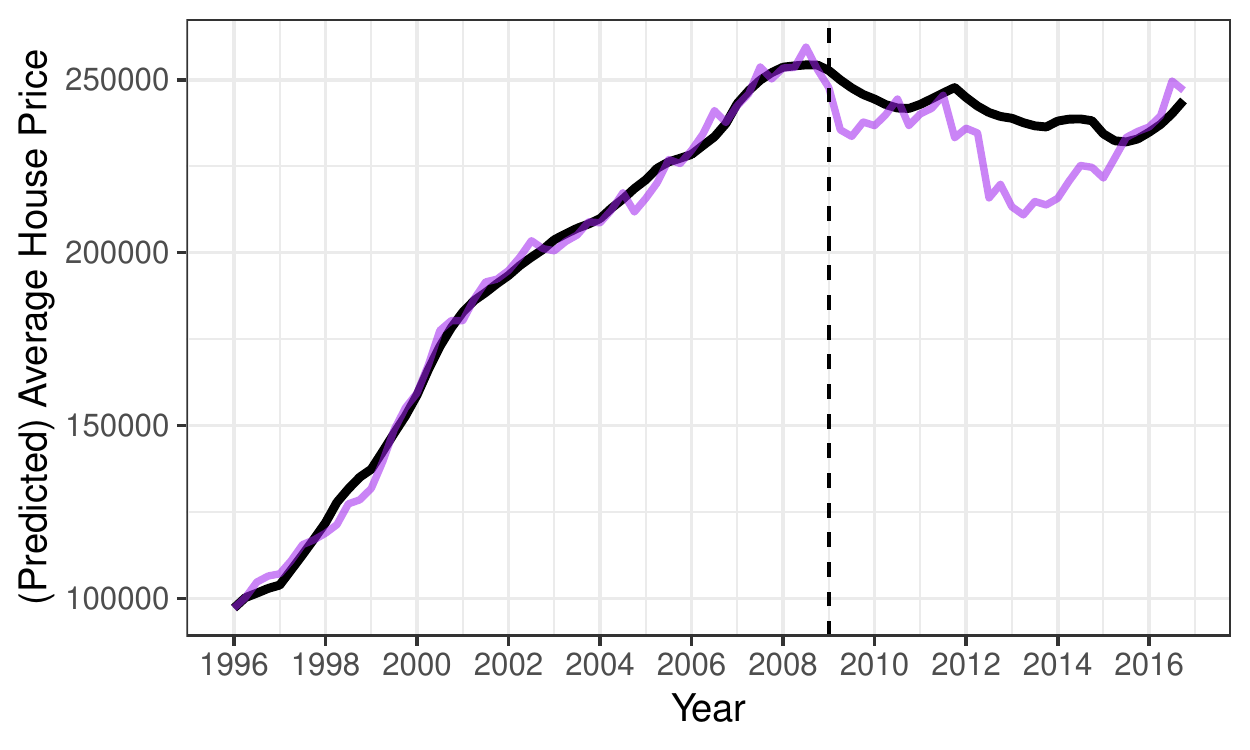}
\endminipage\hfill
\caption{Model Performance(ECM).  Light Purple: Average House Price. Black: Predicted Average House Price.} \label{fig:predict_hp}
\end{figure}

\end{document}